\documentclass[prl,aps,preprint,groupedaddress,showpacs]{revtex4}

\usepackage{pxfonts}
\usepackage{graphicx}
\usepackage{color}
\begin{document}

\title{Kinetics of the inner ring in the exciton emission pattern in GaAs coupled quantum wells}

\author{A.\,T. Hammack}
\author{L.\,V. Butov}
\affiliation{Department of Physics, University of California at San Diego, La Jolla, CA 92093-0319}

\author{J. Wilkes}
\author{L. Mouchliadis}
\author{E.\,A. Muljarov}
\author{A.\,L. Ivanov}
\affiliation{Department of Physics and Astronomy, Cardiff University, Cardiff CF24 3AA, United Kingdom}

\author{A.\,C. Gossard}
\affiliation{Materials Department, University of California at Santa Barbara, Santa Barbara, California 93106-5050}

\begin{abstract}
We report on the kinetics of the inner ring in the exciton emission pattern. The formation time of the inner ring following the onset of the laser excitation is found to be about 30\,ns. The inner ring was also found to disappear within 4\,ns after the laser termination. The latter process is accompanied by a jump in the photoluminescence (PL) intensity. The spatial dependence of the PL-jump indicates that the excitons outside of the region of laser excitation, including the inner ring region, are efficiently cooled to the lattice temperature even during the laser excitation. The ring formation and disappearance are explained in terms of exciton transport and cooling.
\end{abstract}

\pacs{73.63.Hs, 78.67.De, 05.30.Jp}

\date{\today}

\maketitle

\section{Introduction}

An {\em indirect exciton} is a bound pair of an electron and a hole confined in spatially separated layers. It can be realized in coupled quantum well (CQW) structures. The reduction of the overlap of the electron and hole wavefunctions, when they are separated into neighboring quantum wells, results in a large enhancement of the lifetime of indirect excitons comparing to that of regular direct excitons in a single quantum well. This increase in lifetime allows the indirect excitons to travel large distances \cite{Hagn1995, Butov1998, Larionov2000, Butov2002, Voros2005, Ivanov2006, Gartner2006, Gartner2007, Vogele2009}, and to cool down to temperatures $T$ well below the onset of quantum degeneracy that occurs at $T \simeq T_{\rm dB} = (2\pi\hbar^2 n_{\rm x})/(M_{\rm x} g k_{\rm B})\simeq 3$\,K for the density per spin state $n_{\rm x}/g = 10^{10}$\,cm$^{-2}$ \cite{Butov2001} (in the CQWs studied the exciton translational mass is $M_{\rm x} \simeq 0.22\,m_0$ and the spin degeneracy factor is $g=4$). Furthermore, the built-in dipole moment of an indirect exciton $e\!\cdot\!d$ allows control of exciton transport by electrode voltages \cite{Hagn1995, Gartner2006, High2008, High2009, Remeika2009} ($d$ is the separation between the electron and hole layers). The combination of long lifetime, large transport distance, efficient cooling, and an ability to control exciton transport makes the indirect excitons a model system for the investigation of in-plane transport of quasi-two-dimensional (quasi-2D) cold Bose gases in solid state materials.

Studies of indirect excitons in CQWs has lead to the finding of a number of phenomena including exciton pattern formation, a review of which can be found in \cite{Butov2004r}. The features of the exciton emission pattern include the inner ring \cite{Butov2002, Ivanov2006}, external ring \cite{Butov2002, Snoke2002, Butov2004, Rapaport2004, Chen2005, Haque2006}, localized bright spots \cite{Butov2002a, Butov2002, Butov2004, Lai2004, Mouchliadis2007}, and macroscopically ordered exciton state \cite{Butov2002, Butov2004, Levitov2005, Yang2006}. In the regular excitation scheme, where excitons are generated in a micron scale focused laser excitation spot, the inner ring forms around the excitation spot. It was discussed in terms of the cooling of indirect excitons during their propagation away from the excitation spot \cite{Butov2002, Ivanov2006}. However, to date, measurements of the spatial kinetics of the inner ring were unavailable. In this paper, we present studies of the spatially and spectrally resolved kinetics of the exciton inner ring. The results show that the exciton inner ring forms and reaches a steady state within the first few tens of nanoseconds of laser excitation, and also disappears within a few nanoseconds after the laser termination. The spatially-temporal behavior of the inner ring is modelled in terms of in-plane exciton transport and cooling towards the phonon bath (cryostat) temperature.

In Sec.\,II, we describe the experimental data and compare them with numerical simulations of the kinetics of the inner ring in the exciton emission pattern. In Sec.\,III, a model of in-plane transport, thermalization, and photoluminescence (PL) of indirect excitons is presented. In Sec.\,IV, we discuss the results. A short summary of the work is given in Sec.\,V.

\section{Experimental data and numerical simulations}

The measurements were performed using time-resolved imaging with 4\,ns time-resolution and 2\,$\mu$m spatial resolution. Excitons were photogenerated by a pulsed laser at 635\,nm with pulse duration of 500\,ns and edge sharpness of $<1$\,ns, operating with a period of 1\,$\mu$s. The period and duty cycle were chosen such that the photoluminesence pattern of indirect excitons was able to reach equilibrium during the laser excitation and to allow for complete decay of the PL of indirect excitons between laser pulses. The laser is focused to a 10\,$\mu$m full width half maximum (FWHM) excitation spot on the CQW sample. The excitation density $P_{\rm ex}$ was chosen to be below that at which the external ring appears in the emission pattern \cite{Butov2002}. A nitrogen-cooled charge coupled device camera (CCD) coupled to a PicoStar HR TauTec intensifier with a time-integration window of $\delta t=4$\,ns was used to acquire spectral Energy--$y$ PL images at varied delay time $t$. The spectral information was captured by placing the time-gated intensifier and CCD after a single grating spectrometer. The spectral diffraction and time-gated imaging combined allow the direct visualization of the evolution of the indirect exciton PL intensity and energy as a function of delay time $t$ [see Fig.~1\,(a)-(c) for the laser onset and Fig.~1\,(d)-(f) for the laser termination]. Experiments were performed at the applied gate voltage 1.2\,V, peak excitation power 150 $\mu$W, and bath temperature 1.4\,K.

The CQW structure used in these experiments contains two $8$\,nm GaAs QWs separated by a $4$\,nm Al$_{0.33}$Ga$_{0.67}$As barrier. The sample was grown by molecular beam epitaxy (details on the CQW structures can be found in Ref.\,\cite{Butov2002}). The effective spacing between the electron and hole layers is given by $d = 11.5$\,nm \cite{Butov1999}.

\begin{figure}
\begin{center}
\includegraphics[width=10cm]{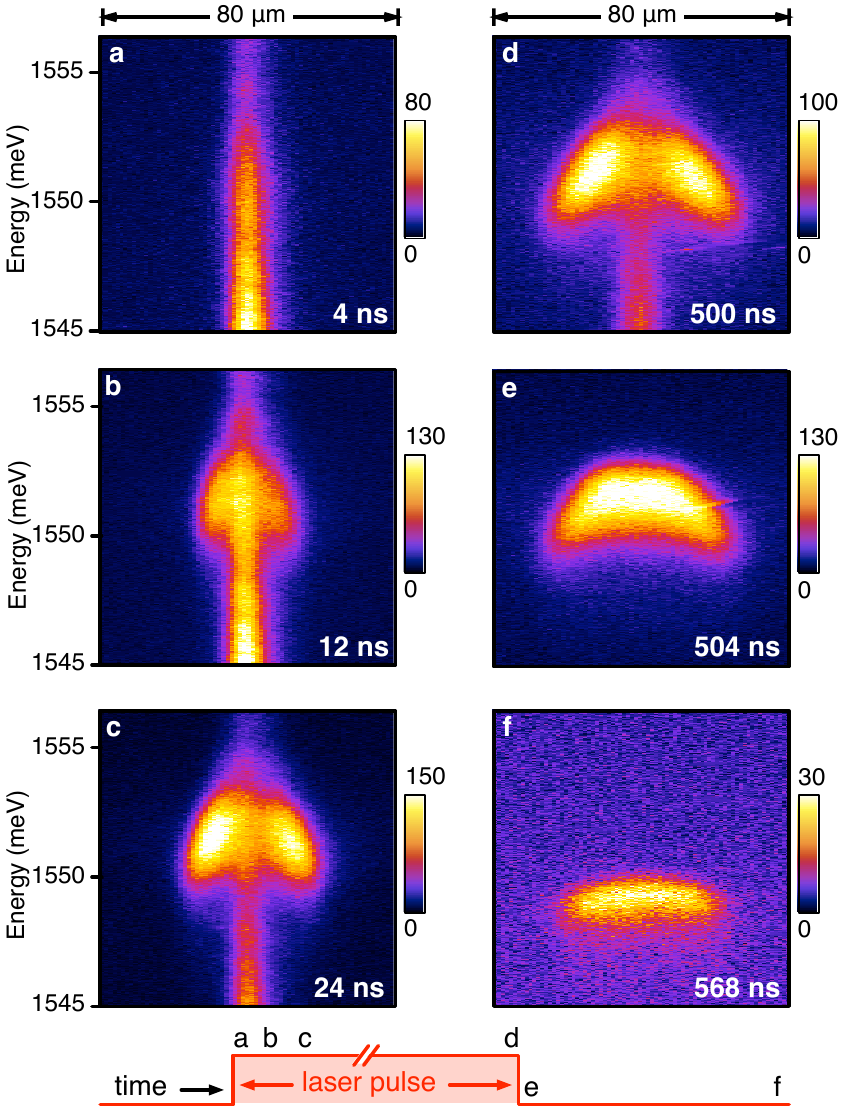}
\caption{The $E-y$ images showing the exciton energy versus radius
during the time evolution of the exciton inner ring following the
onset and termination of the rectangular laser pulse. Time $t=0$\,ns corresponds to the onset of the laser pulse of the duration $\tau_{\rm pulse} = 500$\,ns. Each image is integrated over a time window of $\delta t = 4$\,ns ending at the times (a)-(c) $t = 4$\,ns, 12\,ns, and 24\,ns after the start of the laser pulse, and (d)-(f) $t - \tau_{\rm pulse} = 0$\,ns, 4\,ns, and 68\,ns after its termination. The laser is focused to a 10\,$\mu$m full width half maximum (FWHM) excitation spot on the CQW sample.}
\end{center}
\end{figure}

Fig.~1\,(a)-(c) shows the emergence in time of an arrow shaped profile of the indirect exciton PL signal plotted in the energy $-$ in-plane $y$ axis, $E - y$, coordinates. The central bright stripe corresponds to the emission of bulk excitons. Its spatial profile essentially corresponds to the laser excitation profile. The emission of indirect excitons is observed at the sides of this stripe beyond the excitation spot, due to the exciton transport. The emission energy drops with increasing distance from the origin $r$ (see Fig.~1). The drop in energy with increasing $r$ corresponds to a decrease in density, as detailed in \cite{Ivanov2006} where the inner ring was studied without time resolution. At early times, when the indirect exciton signal is small, the bulk emission, consisting of the central bright stripe, dominates [see Fig.~1\,(a)]. After sufficient time, the exciton inner ring becomes apparent by the presence of a dip in the PL of indirect excitons within the region of laser excitation at the center of the exciton cloud. It is worth noting that the decrease in the exciton PL does not correspond to a dip in the exciton density, which has its maximum at the center of the laser excitation spot (see Fig.~1).

The kinetics presented in Fig.~2\,(a) and (c) show the total spectrally integrated PL intensity of indirect excitons taken from a series of time-gated spectrally resolved images. The data shows that the inner ring forms and reaches a steady state within the first few tens of ns of the laser excitation pulse. Both the spatial and temporal character of the experimental data is in agreement with simulations using a kinetic model for the indirect exciton transport, cooling and optical decay [see Fig.~2\,(a) {\it vs.} 2\,(b), and Fig.~2\,(c) {\it vs.} 2\,(d)]. The model is detailed in Section~III.

\begin{figure}[t]
\begin{center}
\includegraphics[width=10cm]{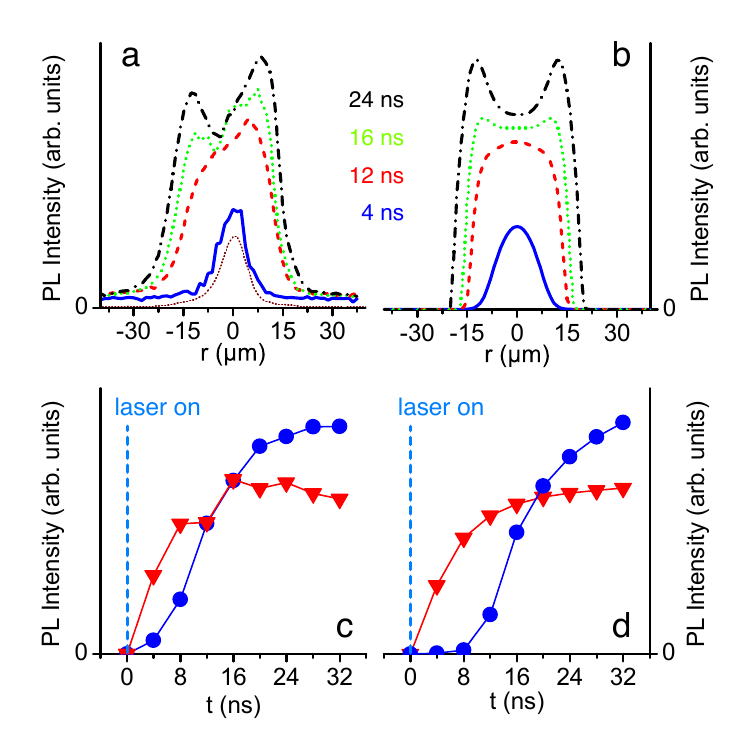}
\caption{Kinetics of the indirect exciton PL profile after the laser excitation onset. The measured (a) and calculated (b) cross-sections of the indirect exciton PL across the diameter of the inner exciton ring as a function of time. The measured (c) and calculated (d) indirect exciton PL intensity at the center of the laser excitation spot ({\color{red}{red $\blacktriangledown$}}) and at the inner ring radius $r = r_{\rm ring} \simeq 12\,\mu$m ({\color{blue}{blue $\medbullet$}}) where the PL maximum signal occurs, as a function of time. The time integration window $\delta t = 4$\,ns for each profile (a), (b) and each point (c), (d). The times $t=0$\,ns and $t=500$\,ns refer to the onset and termination of the rectangular laser excitation pulse. The laser excitation profile is shown by the thin dotted line in (a).
}
\end{center}
\end{figure}

These results demonstrate that within few tens of ns the excitons are able to propagate tens of microns away from the generation region. The large-scale transport is indicative that excitons are capable to screen effectively the disorder potential intrinsic to the quantum wells, leading to enhanced drift and diffusion \cite{Ivanov2002}. This is consistent with the exciton diffusion coefficient $D_{\rm x}$ evaluated with the thermionic model and plotted in Fig.~5\,(d): An increase of $D_{\rm x}$ at a given radius $r$ with increasing time (increasing exciton density) and a decrease of $D_{\rm x}$ at a given time $t$ with increasing radius (decreasing density) are seen. Only low-energy excitons from the radiative zone with kinetic energy $E \leq E_{\gamma} \simeq (E_{\rm x}^2 \varepsilon_{\rm b}) /(2mc^2)$ are optically active \cite{Feldmann1987, Hanamura1988, Andreani1991}, with $E_{\rm x}$ the exciton energy and $\varepsilon_{\rm b}$ the background dielectric constant. The excitons traveling away from the excitation spot cool down towards the lattice temperature. This results in the increase of the occupation of the radiative zone, giving rise to an increase of the emission intensity and therefore leading to the formation of the PL ring. The underlying physics is further illustrated in Fig.\,5, where numerical simulations of $T = T(t)$, $D_{\rm x} = D_{\rm x}(t)$, $n_{\rm x} = n_{\rm x}(t)$, and the optical lifetime of indirect excitons, $\tau_{\rm opt} = \tau_{\rm opt}(t)$, are plotted for the onset of the laser excitation.

Upon termination of the laser pulse [see Fig.~1\,(d)-(f)], an abrupt increase of the PL intensity is detected at the laser excitation spot [see Fig.~3\,(a), (c), and (e)]. After the laser switches off, the optically-dark, high-energy excitons relax to the radiative zone leading to the observed {\it PL-jump}. Experiments performed without spatial resolution have already revealed the PL-jump \cite{Butov2001}. However, the results of the time-resolved imaging experiments presented here clarify that the PL-jump is observed predominantly within the laser excitation spot, where indirect excitons are heated by the laser. Within 4\,ns, the time resolution of the current experiments, the excitons cool down to the lattice temperature $T_{\rm b}=1.4$\,K. The characteristic cooling (thermalization) time, as calculated with the model described in Section~III, is $\tau_{\rm th} \simeq 0.2$\,ns [see Fig.~5\,(a)]. The contrast of the PL-jump is defined as $(I_{\rm max} - I_{\rm laser\,on})/I_{\rm laser\,on}$ with $I_{\rm max} = I_{\rm max}(r)$ and $I_{\rm laser\,on} = I_{\rm laser\,on}(r)$ the maximum PL intensity after the laser pulse termination and the steady-state PL-intensity in the presence of the laser pulse, respectively. The measured contrast of the PL-jump against the radial distance $r$ is plotted in Fig.~3\,(c). Averaging the numerical simulations [see inset in 3\,(d)] over the 4\,ns integration window to match the experimental conditions leads to the PL-jump contrast shown in Fig.~3\,(d), in agreement with the experiment [see Fig.~3\,(c) and (e) {\it vs}. Fig.~3\,(d) and (f)].

\begin{figure}
\begin{center}
\includegraphics[width=9cm]{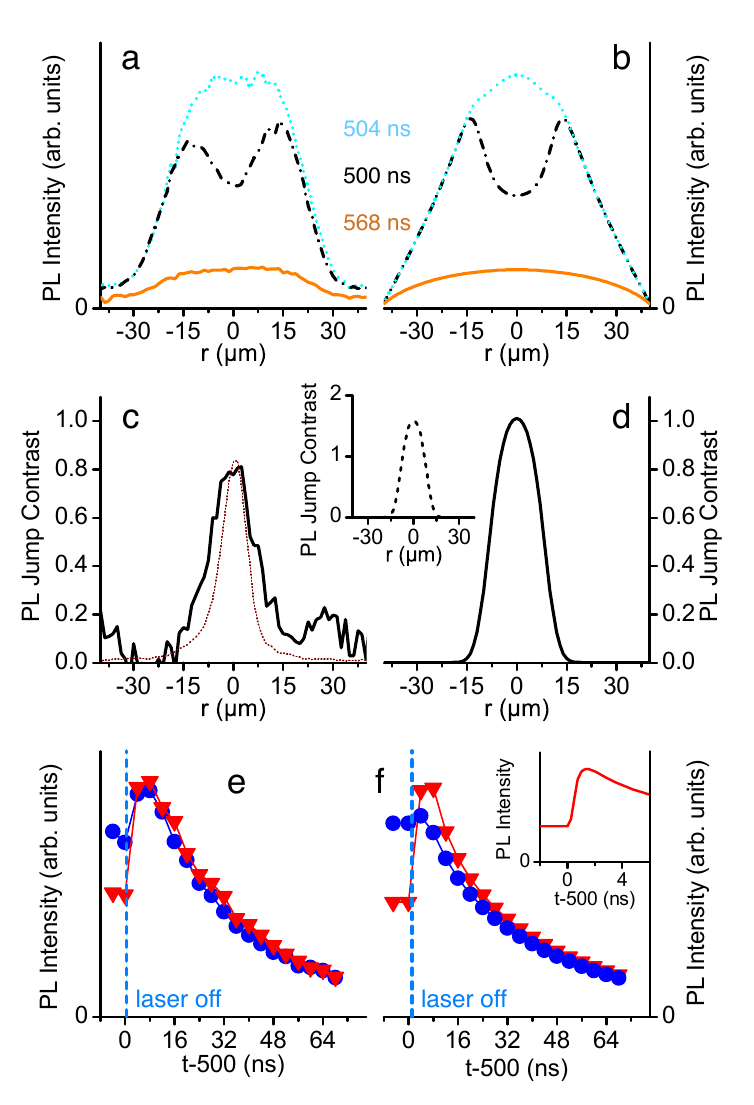}
\caption{Kinetics of the indirect exciton PL profile during the laser excitation termination. The measured (a) and calculated with Eqs.\,(\ref{diff})-(\ref{pl}) (b) spatial profiles of the PL signal from indirect excitons across the inner ring at the times $t = 500$\,ns (dash-dotted line), 504\,ns (dotted line), and 568\,ns (solid line). The times $t=0$\,ns and $t=500$\,ns refer to the onset and termination of the rectangular laser excitation pulse. The measured (c) and evaluated numerically (d) contrast of the PL-jump $(I_{\rm max} - I_{\rm laser\,on})/I_{\rm laser\,on}$ against the radial coordinate. The laser excitation profile is shown by the dotted line in (c). The measured (e) and calculated (f) PL intensity at the center of the laser excitation spot ({\color{red}{red $\blacktriangledown$}}) and at the radial distance where the PL maximum intensity occurs, $r = 12\,\mu$m, ({\color{blue}{blue $\medbullet$}}) as a function of time. Insets: The contrast of the PL-jump (d) and the PL-jump at the center of excitation (f), evaluated with Eqs.\,(\ref{diff})-(\ref{pl}) without time integration to match 4\,ns experimental resolution. Apart from the insets, each calculated curve is smoothed by the device resolution function with the time integration window $\delta t = 4$\,ns to match the experimental conditions.}
\end{center}
\end{figure}

The monotonic decrease of the PL-jump with increasing radius [see Fig.~3\,(c)-(d)] demonstrates that the effective exciton temperature $T$ during the laser excitation lowers with increasing $r$. This is consistent with the model we use: The numerical simulations of the exciton temperature profile, plotted in the inset of Fig.~6\,(a) for two time delays 0\,ns and 4\,ns after the termination of the laser pulse are in agreement with the spatial dependence of the PL-jump shown in Fig.~3. Both the experimental data and calculations demonstrate that the exciton cooling time to the lattice temperature is much shorter than the exciton lifetime $\tau_{\rm opt} \simeq 50$\,ns.

Both the excitation and emission patterns (a laser excitation spot and a PL-ring around the excitation spot, which develops at large delay times), studied in the present work, correspond to the geometry inverted comparing to that used for the optically induced traps (a ring-shaped laser excitation and a PL pattern concentrated at the center of the excitation ring, which builds up at large delay times) \cite{Hammack2006}. The kinetics of the inner PL-ring and the optically induced trap are quantitatively consistent with each other.

\section{Model}

A set of coupled nonlinear differential equations we use in order to model transport, thermalization, and photoluminescence of indirect excitons is given by
\begin{eqnarray}
\frac{\partial n_{\rm x}}{\partial t} &=& \nabla \Big[ D_{\rm x}
\nabla n_{\rm x} + \mu_{\rm x} n_{\rm x} \nabla \left( u_0 n_{\rm
x} + U_{\rm QW} \right) \Big] - \Gamma_{\rm opt} n_{\rm x} +
\Lambda^{\rm (x)} \, ,
\label{diff} \\
\frac{\partial T}{\partial t}  &=& \left( \frac{\partial
T}{\partial t} \right)_{n_{\rm x}} + \ S_{\rm pump} + S_{\rm opt}
\,, \ \ \ \mbox{where}
\nonumber \\
&&\left( {\partial T \over \partial t} \right)_{n_{\rm x}}
\!\!\!\! = - { 2 \pi \over \tau_{\rm sc} } \left( {T^2 \over
T_{\rm dB}} \right) \big(1 - e^{-T_{\rm dB}/T} \big) \!\!
\int_1^{\infty} \!\!\! d \varepsilon \ \varepsilon
\sqrt{\varepsilon \over \varepsilon - 1}
\nonumber \\
&& \ \ \ \ \ \ \ \ \ \ \ \ \ \ \ \ \ \ \ \ \ \ \ \ \times
\frac{\left|F_z \left(a \sqrt{\varepsilon(\varepsilon - 1)}\right)
\right|^2}{(e^{\varepsilon E_0 / k_{\rm B} T_{\rm b}} - 1) } \
{e^{\varepsilon E_0 / k_{\rm B} T_{\rm b}} - e^{\varepsilon E_0 /
k_{\rm B} T} \over (e^{\varepsilon E_0 / k_{\rm B} T} + e^{-T_{\rm
dB}/T} - 1) } \, ,
\label{therm} \\
I^{\rm sig}_{\rm PL} &=& \Gamma_{\rm opt}^{\rm sig} \, n_{\rm
x}\,, \ \ \ \mbox{where}
\nonumber \\
&&\Gamma_{\rm opt}^{\rm sig} = { 1 \over 2 \tau_{\rm R}} \Bigg(
{E_{\gamma} \over k_{\rm B} T_{\rm dB} } \Bigg) \int_{z_{\rm
s}}^{1} \!\! { 1 + z^2 \over [(e^{E_{\gamma}/k_{\rm B} T}) / (1 -
e^{-T_{\rm dB}/T})] e^{- z^2 E_{\gamma}/k_{\rm B} T} - 1 }\,\,dz
\, . \label{pl}
\end{eqnarray}
Equations (\ref{diff})-(\ref{pl}) describe in-plane profiles of the density $n_{\rm x} = n_{\rm x}(r,t)$, effective temperature $T = T(r,t)$ and signal PL intensity $I^{\rm sig}_{\rm PL} = I^{\rm sig}_{\rm PL}(r,t)$ of indirect excitons.

In the drift-diffusion Eq.\,(\ref{therm}) for in-plane transport of the particles \cite{Ivanov2002}, $D_{\rm x}$, $\mu_{\rm x}$, $\Gamma_{\rm opt}$, and $\Lambda^{\rm (x)}$ are the diffusion coefficient, mobility, radiative decay rate, and generation rate of indirect excitons, respectively. The $\nabla$-operator has only the radial component, $\nabla_r = \partial/\partial r$, due to the cylindrical symmetry one deals with. The mobility $\mu_{\rm x}$ is given in terms of the diffusion coefficient $D_{\rm x}$ through the generalized Einstein relationship, $\mu_{\rm x} = D_{\rm x} [(e^{T_{\rm dB}/T} - 1)/(k_{\rm B}T_{\rm dB})]$ \cite{Ivanov2002}. The random potential $U_{\rm QW} = U_{\rm rand}({\bf r})$ on the right-hand side (r.h.s.) of Eq.\,(\ref{diff}) is mainly due to the CQWs thickness and alloy fluctuations. The drift term $\propto \nabla (u_0 n_{\rm x} + U_{\rm QW})$ stems from the dipole-dipole interaction of indirect excitons and the in-plane potential $U_{\rm QW}$. For the first contribution we use $u_0 = 4 \pi d (\mbox{e}^2/\varepsilon_{\rm b})$. This corresponds to the mean-field approximation of the interaction energy of indirect excitons. The latter stems from the dipole-dipole repulsion between the particles and gives rise to the blue shift of the PL line. For $n_{\rm x} \gtrsim 10^9\,\mbox{cm}^{-2}$ relevant to the experiment, the correlation energy of exciton-exciton interaction is less than the mean-field energy \cite{Ivanov2009} and therefore is neglected in the present model. The radiative rate $\Gamma_{\rm opt} = 1/\tau_{\rm opt}$ on the r.h.s. of Eq.\,(\ref{diff}) is $\Gamma_{\rm opt} = \Gamma_{\rm opt}^{\rm sig}(z_{\rm s}\!=\!0)$ with $\Gamma_{\rm opt}^{\rm sig}$ given by Eq.\,(\ref{pl}).

The first term on the r.h.s. of Eq.\,(\ref{therm}), $(\partial T / \partial t)_{n_{\rm x}}$, describes thermalization (cooling) of indirect excitons, due to their interaction with a bath of bulk acoustic phonons at temperature $T_{\rm b}$ \cite{Ivanov1999}. Here, $\tau_{\rm sc} = (\pi^2 \hbar^4 \rho)/(D_{\rm dp}^2 M_{\rm x}^3 v_{\rm LA})$ is the characteristic scattering time, $E_0 = 2 M_{\rm x} v_{\rm LA}^2$ is the characteristic energy of the longitudinal acoustic (LA) phonon assisted thermalization at low temperatures, $v_{\rm LA}$ is the velocity of long-wave-length LA phonons, $\rho$ is the crystal density, and $D_{\rm dp} = D_{\rm c} - D_{\rm v}$ is the deformation potential of exciton -- LA-phonon interaction. The form-factor $F_z(a \sqrt{\varepsilon(\varepsilon - 1}))$ refers to a rigid-wall confinement potential of quantum wells, where $a = (d_{\rm QW} M_{\rm x} v_{\rm LA})/\hbar$ with $d_{\rm QW}$ the quantum well thickness and $\varepsilon = E/E_0$ the normalized single-particle kinetic energy of excitons. The terms $S_{\rm pump}$ and $S_{\rm opt}$ on the r.h.s. of Eq.\,(\ref{therm}), which are detailed in Ref.\,\cite{Ivanov2004}, deal with heating of indirect excitons by the laser pulse and recombination heating or cooling of the particles. The laser-induced heating is given by
\begin{equation}
S_{\rm pump} = \frac{ E_{\rm inc} - k_{\rm B} T I_2 }{ 2 k_{\rm B}
T I_1 - k_{\rm B} T_{\rm dB} I_2 } \, \Lambda_{T_{\rm dB}}^{\rm
(x)} \,, \label{laser}
\end{equation}
where $\Lambda_{T_{\rm dB}}^{\rm (x)} = [(\pi \hbar^2)/ (2 k_{\rm B} M_{\rm x})] \Lambda^{\rm (x)}(r,t)$ and $E_{\rm inc} \gg k_{\rm B}T_{\rm b}$ is an average kinetic energy of high-energy indirect excitons injected into the CQW structure by means of photocarriers. The latter are generated in the GaAs and AlGaAs layers by the laser pulse. The term $S_{\rm opt}$, which takes into account a contribution from the ``optical evaporation'' of low-energy indirect excitons to the total energy balance \cite{Ivanov2004}, is determined by
\begin{equation}
S_{\rm opt} = \frac{ k_{\rm B}T I_2 \Gamma_{\rm opt} - E_{\gamma}
\Gamma_{\rm opt}^{\rm E} }{ 2 k_{\rm B} T I_1 - k_{\rm B} T_{\rm
dB} I_2 } \, T_{\rm dB} \, . \label{evaporation}
\end{equation}
Here, the energy rate $\Gamma_{\rm opt}^{\rm E}$ , due to the optical decay, is given by
\begin{equation}
\Gamma_{\rm opt}^{\rm E} = { 1 \over 2 \tau_{\rm R}} \Bigg(
{E_{\gamma} \over k_{\rm B} T_{\rm dB} } \Bigg) \int_{0}^{1} \!\!
{ 1 - z^4 \over [(e^{E_{\gamma}/k_{\rm B} T}) / (1 - e^{-T_{\rm
dB}/T})] e^{- z^2 E_{\gamma}/k_{\rm B} T} - 1 }\,\,dz \, .
\label{optE}
\end{equation}
In Eqs.\,(\ref{laser}) and (\ref{evaporation}), the parameters
$I_{1,2} = I_{1,2}(T_{\rm dB}/T)$ are $I_1 = (1 - e^{-T_{\rm
dB}/T}) \int_0^{\infty} dz [z/(e^z + e^{-T_{\rm dB}/T} - 1)]$ and
$I_2 = e^{-T_{\rm dB}/T} \int_0^{\infty} dz [(z e^z)/(e^z +
e^{-T_{\rm dB}/T} - 1)^2]$.

Finally, the intensity $I^{\rm sig}_{\rm PL}$ of the PL signal, collected in the normal direction within an aperture angle $\alpha$ (in the experiment, $\alpha \simeq 30^{\circ}$), is given by Eq.\,(\ref{pl}). In this case, the lower integration limit in the expression for the decay rate $\Gamma_{\rm opt}^{\rm sig}$ is $z_{\rm s} = 1 - \sin^2(\alpha/2)$ [see Eq.\,(\ref{pl})]. Both $\Gamma_{\rm opt}^{\rm sig}$ and $\Gamma_{\rm opt}$ are inversely proportional to the intrinsic radiative lifetime $\tau_{\rm R}$ of the exciton ground-state with zero in-plane momentum.

In order to evaluate the random drift term $\mu_{\rm x} n_{\rm x} \nabla( U_{\rm QW})$ on the r.h.s. of the drift-diffusion Eq.\,(\ref{diff}), we implement a thermionic model \cite{Ivanov2002,Ivanov2006}. In this approach, the influence of disorder is approximately taken into account by using the disorder-dependent effective diffusion coefficient:
\begin{equation}
D_{\rm x} = D_{\rm x}^{(0)} \exp \bigg[ - \frac{U^{(0)}}{k_{\rm
B}T + u_0 n_{\rm x}} \bigg] \,, \label{thermionic}
\end{equation}
where $D_{\rm x}^{(0)}$ is the input diffusion coefficient in the absence of CQW disorder, and $U^{(0)}/2 = \langle |U_{\rm rand}({\bf r}) - \langle U_{\rm rand}({\bf r}) \rangle|\rangle$ is the amplitude of the disorder potential. Equation~(\ref{thermionic}) describes the temperature and density dependent screening of the long-range-correlated disorder potential $U_{\rm QW} = U_{\rm rand}({\bf r})$ by dipole-dipole interacting indirect excitons. The vanishing screening at the external edge of the inner PL ring, due to reducing exciton density, leads to a strong suppression of the exciton propagation away from the excitation spot and, as a result, to the sharp contrast of the ring \cite{Ivanov2006}.

There are two main features in our present numerical simulations comparing to those reported earlier \cite{Ivanov2006} in order to model a steady-state inner PL ring: (i) For the first time we have performed high-resolution numerical simulations with Eqs.\,(\ref{diff})-(\ref{pl}) in a space-time domain, and (ii) in order to mimic more closely the experiment, we model the source term $\Lambda^{\rm (x)}$ of indirect excitons [see the r.h.s. of Eq.\,(\ref{diff})] by assuming a generation of incoming indirect excitons, as secondary particles, from laser-induced
photocarriers.

In order to express $\Lambda^{\rm (x)}$ via the generation rate $\Lambda^{\rm (0)}$ of free electron - hole pairs, which are photoexcited in the cladding AlGaAs layers and captured by the GaAs CQW structure, we implement the quantum mass action law (QMAL). According to the QMAL, a total number of electron-hole pairs is distributed among the bound (exciton) and unbound states. For quasi-2D indirect excitons, the QMAL reads as
\cite{Reinholz2002,Mouchliadis2007}
\begin{equation}
n_{\rm x} = - {2 M_{\rm x} k_{\rm B}T \over \pi \hbar^2} \ln \Big[
1 - e^{\epsilon_{\rm x}/(k_{\rm B} T)} \big( e^{T_{\rm dB}^{\rm
e}/T} - 1 \big) \big( e^{T_{\rm dB}^{\rm h}/T} - 1 \big) \Big] \ ,
\label{QMAL}
\end{equation}
where $\epsilon_{\rm x}$ is the (indirect) exciton binding energy, and the electron (hole) quantum degeneracy temperature is given by $k_{\rm B} T_{\rm dB}^{\rm e(h)} = \big[(\pi \hbar^2)/m_{\rm e(h)}\big]n_{\rm e(h)}$ with $m_{\rm e(h)}$ and $n_{\rm e(h)}$ the electron (hole) mass and concentration, respectively. Equation (\ref{QMAL}) characterizes a quasi-equilibrium balance between $n_{\rm x}$ and $n_{\rm e} = n_{\rm h}$. A typical time $\tau_{\rm QMAL}$ needed to quasi-equilibrate the system of electrons, holes and indirect excitons is comparable to that of binding of photoexcited electrons and holes in excitons. The latter one is about $10-30$\,ps for $n_{\rm x} \simeq 10^{10}\,\mbox{cm}^{-2}$ and helium temperatures \cite{Damen1990, Strobel1991, Blom1993, Gulia1997, Szczytko2004}. Because $\tau_{\rm QMAL}$ is much less than the characteristic times of the thermalization and transport processes, $\tau_{\rm th} \sim 0.1$\,ns and $\tau_{\rm diff} \sim 1$\,ns, the use of the QMAL is justified. For the case $T_{\rm dB}^{\rm e(h)} \ll T$, relevant to the experiment, Eq.\,(\ref{QMAL}) yields:
\begin{equation}
\Lambda^{\rm (x)} = \Lambda^{\rm (0)} \ \frac{ 4 M_{\rm x} T_{\rm
dB}^{1/2} }{ (m_{\rm e} m_{\rm h} T)^{1/2} e^{-\epsilon_{\rm x}/(2
k_{\rm B} T)} + 4 M_{\rm x} T_{\rm dB}^{1/2} } \ . \label{rate}
\end{equation}
Note that in Eq.\,(\ref{rate}) the degeneracy temperature $T_{\rm dB}$ is proportional to the accumulated density of indirect excitons, $n_{\rm x} = n_{\rm x}(r,t)$. According to numerical evaluations of Eq.\,((\ref{rate}) adapted to the experimental conditions, apart from the first few hundred picoseconds after the onset of the laser excitation, when $T \gtrsim 10$\,K and $n_{\rm x} \lesssim 10^{9}\,\mbox{cm}^{-2}$, one has $n_{\rm x} \gg n_{\rm e(h)}$, see Fig.\,4, and $\Lambda^{\rm (x)} \simeq \Lambda^{\rm (0)}$. Formally, this is because in the denominator on the r.h.s. of Eq.\,(\ref{rate}) the term $(m_{\rm e} m_{\rm h} T)^{1/2} e^{-\epsilon_{\rm x}/(2 k_{\rm B} T)}$ is much less than $4 M_{\rm x} T_{\rm dB}^{1/2}$. In this case, injected electron-hole pairs very effectively transfer to the exciton system.

\begin{figure}
\begin{center}
\includegraphics[width=10cm]{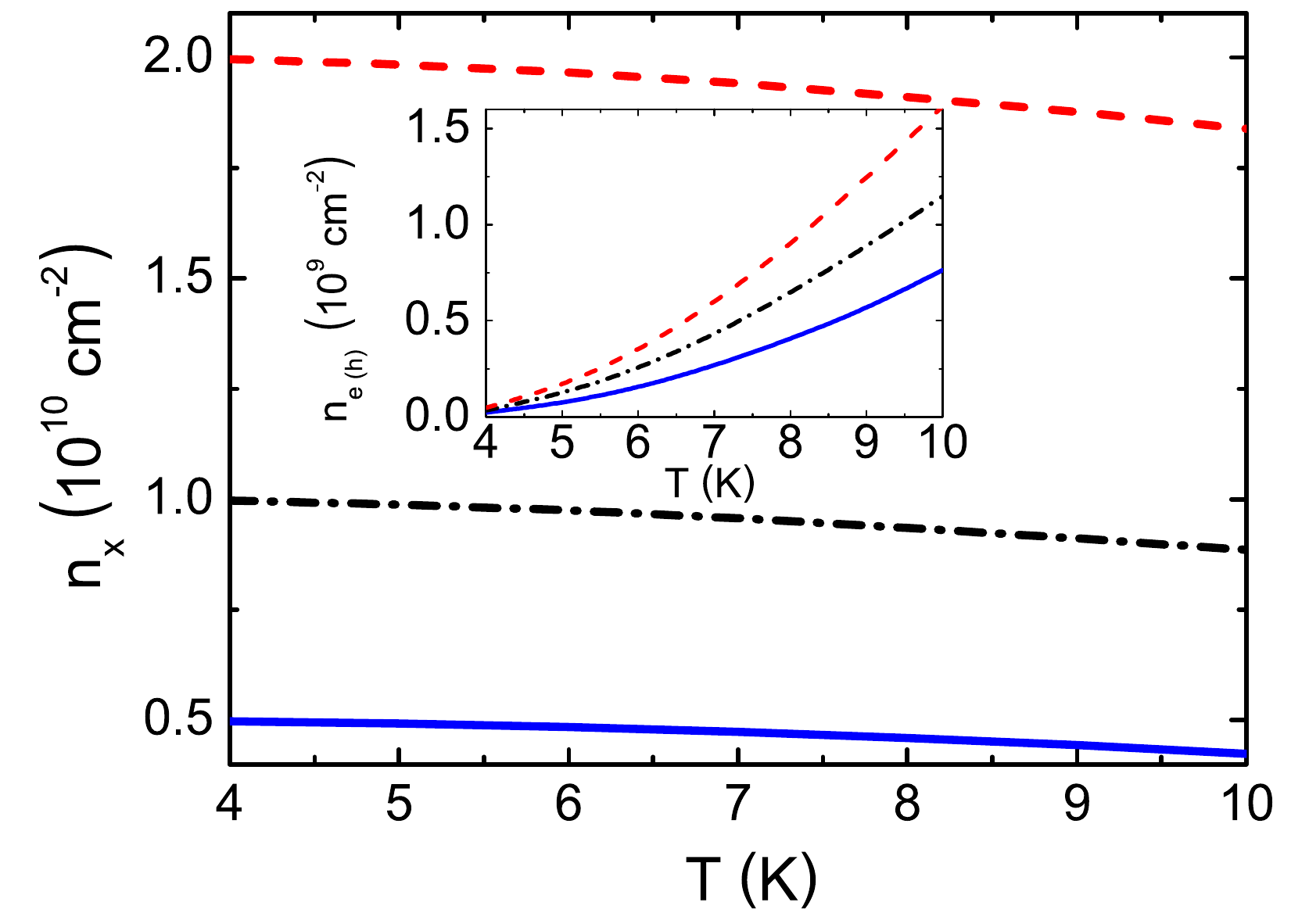}
\caption{Exciton and electron (hole) concentrations, $n_{\rm x}$ and $n_{\rm e} = n_{\rm h}$, as a function of temperature, evaluated by using the quantum mass action law: $n_{\rm x} + n_{\rm e(h)} = 2 \times 10^{10}\,\mbox{cm}^{-2}$ (dashed lines), $10^{10}\,\mbox{cm}^{-2}$ (dash-dotted lines), and $0.5 \times 10^{10}\,\mbox{cm}^{-2}$ (solid lines).}
\end{center}
\end{figure}

The observed inner PL ring is nearly radially symmetric in space, when taking the center of the excitation spot as the origin. Thus in numerical simulations with Eqs.\,(\ref{diff})-(\ref{pl}), a polar coordinate system is used and the condition invokes that none of the quantities modelled have any angular dependence. According to the spatial profile of the laser pulse, we assume that the generation rate $\Lambda^{\rm (0)}$ is given by the Gaussian:
\begin{equation}
\Lambda^{\rm (0)} = \Lambda^{\rm (0)}(r,t) = \Lambda^{\rm
(0)}(r\!=\!0,t) \, e^{-r^2/r_0^2}, \label{gauss}
\end{equation}
with $r_0$ the radius of the excitation spot [$r_0 = \mbox{FWHM}/(2 \sqrt{\ln2}) = 5.8\,\mu$m]. The temporal shape of the rectangular laser pulse with the Gaussian edges is modelled by $\Lambda^{\rm (0)}(r\!=\!0,t)$, where $\Lambda^{\rm (0)}(r\!=\!0,t) = \tilde{\Lambda}^{(0)} \,\exp[-\sigma (t - t_0)^2]$ for $t \leq t_0 = 0.6\,\mbox{ns}$, $\tilde{\Lambda}^{(0)}$ for $t _0 \leq t \leq \tau_{\rm pulse} = 500\,\mbox{ns}$, and $\tilde{\Lambda}^{(0)} \,\exp[-\sigma (t - \tau_{\rm pulse})^2]$ for $t \geq \tau_{\rm pulse}$, and $\sigma = 15.3\,\mbox{ns}^{-2}$.

\begin{figure}
\begin{center}
\includegraphics[width=12cm]{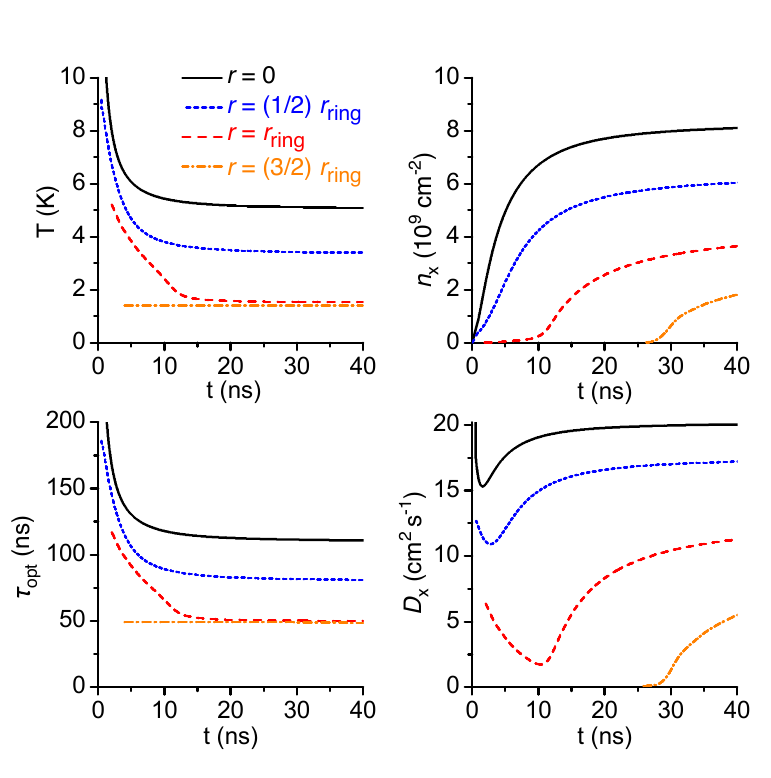}
\caption{Modelling of the transient dynamics of the exciton temperature $T$ (a), density $n_{\rm x}$ (b), optical lifetime $\tau_{\rm opt}$ (c), and in-plane diffusion coefficient $D_{\rm x}$ (d) for the onset of the laser excitation. The radial coordinate is $r = 0$ (solid line), $r_{\rm ring}/2$ (dotted line), $r_{\rm ring}$ (dashed line), and $3r_{\rm ring}/2$ (dash-dotted line).}
\end{center}
\end{figure}

An explicit finite difference scheme (EFDS) is used to integrate the drift-diffusion Eq.\,(\ref{diff}), the only one differential equation among Eqs.\,(\ref{diff})-(\ref{pl}) that includes both derivatives, with respect to $t$ and $r$. In numerical simulations, the time and radial coordinate steps are chosen to be $\Delta t = 6.25$\,ps and $\Delta r = 1\,\mu$m, respectively. In this case the dimensionless parameter $D_{\rm x} \Delta t / (\Delta r)^2$ is about 0.01875, insuring high stability of the EFDS. The various integrals relevant to Eqs.\,(\ref{therm})-(\ref{optE}) are calculated using an adaptive Simpson algorithm. Here, special care is taken to evaluate integrands correctly in the limits of low temperature and high density.

\begin{figure}
\begin{center}
\includegraphics[width=12cm]{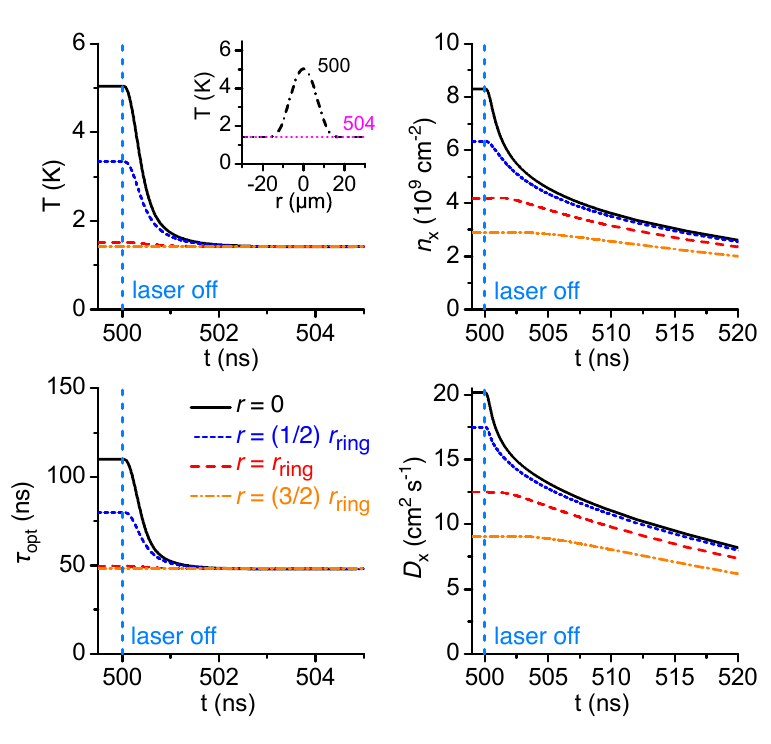}
\caption{Modelling of the transient dynamics of the exciton temperature $T$ (a), density $n_{\rm x}$ (b), optical lifetime $\tau_{\rm opt}$ (c), and in-plane diffusion coefficient $D_{\rm x}$ (d) for the termination of the laser excitation. The radial coordinate is $r = 0$ (solid line), $r_{\rm ring}/2$ (dotted line), $r_{\rm ring}$ (dashed line), and $3r_{\rm ring}/2$ (dash-dotted line). Inset: The calculated spatial profile of the exciton temperature, $T = T(r)$, for $t = 500$\,ns (solid line) and $t = 504$\,ns (dotted line).}
\end{center}
\end{figure}

In order to clarify further the underlying physics of the inner PL ring, in Figs.~5 and 6 we plot $T = T(t)$, $n_{\rm x} = n_{\rm x}(t)$, $\tau_{\rm opt} = \tau_{\rm opt}(t)$, and $D_{\rm x} = D_{\rm x}(t)$, modelled for the onset and termination edges of the laser excitation pulse, respectively. The shown dynamics refer to the center of the excitation spot (solid lines), the radial distance $r = r_{\rm ring}$ where a maximum PL signal occurs (dashed lines), $r = r_{\rm ring}/2$ (dotted lines), and $r = 3r_{\rm ring}/2$ (dash-dotted lines).

For the onset of the laser excitation, Figs.~5\,(a) and (b) illustrate the thermalization kinetics of indirect excitons and gradual building up of the density, respectively. The monotonous decrease of $\tau_{\rm opt}$ with time is due to the cooling of indirect excitons [see Fig.~5\,(c)]. A steady-state value of the optical lifetime decreases with increasing $r$, because the laser induced heating of the exciton system decreases with the radial distance from the laser spot center. The initial rapid decrease of $D_{\rm x}$ [see Fig.~5\,(d)] originates from the thermalization of the
exciton system: Cold excitons cannot overcome the in-plane disorder potential and become localized. A further increase of the diffusion coefficient after the thermalization transient is due to the screening effect which develops with increasing $n_{\rm x}$.

For the termination edge of the laser pulse, the decay dynamics shown in Fig.~6 are dominated by the two characteristic times, thermalization time $\tau_{\rm th}$ and optical decay time $\tau_{\rm opt} \gg \tau_{\rm th}$. Within first $1-2$\,ns the complete thermalization occurs [see Fig.~6\,(a)], and the system of indirect excitons decays with $\tau_{\rm opt} = \tau_{\rm opt}(T\!=\!T_{\rm b}) \simeq 50$\,ns [see Figs.~6\,(b) and (c)]. The gradual decrease of $D_{\rm x}$ at delay times much larger
than $\tau_{\rm th}$ [see Fig.~6\,(d)] is due to the relaxation of the screening effect with decreasing density of indirect excitons.

\section{Discussion}

Numerical simulations with Eqs.\,(\ref{diff})-(\ref{pl}) quantitatively reproduce the experimental data [see in Figs.\,2 and 3 experiment~{\it vs.}~modelling] for the following control parameters: $U_0 = 0.7$\,meV, $D_{\rm x}^{(0)} = 30\,\mbox{cm}^2/\mbox{s}$, $\tilde{\Lambda}^{(0)} = 2 \times 10^9\,\mbox{cm}^{-2} \mbox{ns}^{-1}$, and $E_{\rm inc} = 12.9$\,meV. Although we use four fitting parameters, the procedure is well justified: (i) The whole set of the experimental data, measured at various $r$ and $t$, are modelled with the same values of the control parameters, and (ii) the fitting parameters influence different aspects of the transport and PL processes in a separate way, i.e., can be inferred independently. The pump rate $\tilde{\Lambda}^{(0)}$ yields a maximum concentration $n_{\rm x}^{\rm max} = n_{\rm x}(r\!=\!0,t\!=\!\tau_{\rm pulse})$, which, in turn, can be evaluated from the blue shift of the PL line. The average energy of incoming, hot indirect excitons, $E_{\rm inc}$, governs the contrast of the PL-jump. The in-put diffusion coefficient $D_{\rm x}^{(0)}$ determines the time-dependent radius of the inner PL ring. Finally, the amplitude $U_0$ of the long-range-correlated disorder potential is responsible for the spatial pinning of the PL signal at ring edges. Note that the above values of the control parameters are consistent with those used in our previous simulations for the steady-state inner ring \cite{Ivanov2006} and laser-induced traps \cite{Hammack2006, Hammack2007} studied for the same CQW structures. Table~I lists the inferred values of the control parameters as well as the known values, the basic parameters, and the parameters of the model.

\begin{table}[ht]
\caption{List of Parameters} \centering
\begin{tabular}{|c c| |c c|}
\hline \multicolumn{2}{|c||}{Basic Parameters} & \multicolumn{2}{|c|}{Model Parameters}\\ [0.5ex] \hline
$M_{\rm x}       $ & $0.22\,{\rm m_0}                    $ & $D_{\rm x}^{(0)}      $ & $30\,{\rm cm^2\, s^{-1}}              $ \\
${\rm \tau_{sc}} $ & $110\,{\rm ns}                      $ & $U_0                  $ & $0.7\,{\rm meV}                       $ \\
${\rm \tau_{R}}  $ & $20\,{\rm ns}                       $ & $\tilde{\Lambda}^{(0)}$ & $2 \times 10^9\,{\rm cm^{-2}\,ns^{-1}}$ \\
${\rm \nu_s}     $ & $3.7\times10^5\,{\rm cm\,s^{-1}}    $ & $E_{\rm inc}          $ & $12.9\,{\rm meV}                      $ \\
$\rho            $ & $5.3\,{\rm g\,cm^{-3}}              $ &  $u_0             $ & $1.6\times 10^{-10}\,{\rm meV\,cm^2}$\\
$D_{\rm dp}      $ & $8.8\,{\rm eV}                      $ &                         &                                         \\
${\rm T_b}       $ & $1.4\,{\rm K}                       $ &                         &                                         \\
$E_0             $ & $34.2\,{\rm \mu eV}                 $ &                         &                                         \\
$E_{\gamma}      $ & $138\,{\rm \mu eV}                  $ &                         &                                         \\
$L_z             $ & $8\,{\rm nm}                        $ &                         &                                         \\
\hline
\end{tabular}
\label{table:parameters}
\end{table}

The model includes the nonclassical, quantum-statistical effects in the description of the transport, thermalization and optical decay of indirect excitons: Equations (\ref{diff})-(\ref{optE}) and (\ref{QMAL})-(\ref{rate}) explicitly depends upon $T_{\rm dB}$. However, the quantum corrections are rather minor, due to relatively weak laser excitations used in the experiment. The quantum effects are not required for the inner ring or the PL-jump formation. Both are classical phenomena associated with the exciton cooling when they travel away from the excitation spot (in the case of the inner ring) or when the excitation pulse is terminated (in the case of the PL jump). However, the quantum degeneracy effects become essential for the dynamics and contrast of the inner PL ring and PL-jump, if smaller bath temperatures or higher excitation powers are used \cite{Butov2001, Ivanov2006, Hammack2006}.
For instance, bosonic stimulation of exciton scattering can lead to the enhancement of the exciton scattering rate to the low-energy optically active states with increasing exciton concentration as described in Ref.\,\cite{Butov2001}. 

Note that a ring in the emission pattern can form both in an exciton system \cite{Butov2002,Ivanov2006} and in an electron-hole plasma (EHP) \cite{Stern2008}. In both cases the requirements for the ring formation in the emission pattern include the long lifetime of the carriers, which allows transport over substantial distances, and cooling of the carriers during their transport away from the excitation spot, which leads to the increase of the emission intensity. However, the exciton system can be distinguished from EHP by the emission linewidth. For a neutral quasi-2D EHP, the emission linewidth should be about the sum of the electron and hole Fermi energies, $\Delta_{\rm EHP} \simeq k_{\rm B} T_{\rm dB}^{\rm e} + k_{\rm B} T_{\rm dB}^{\rm h} = \pi \hbar^2 n_{\rm e(h)} (1/m_{\rm e} + 1/m_{\rm h})$, with $n_{\rm e} = n_{\rm h}$ the density of electrons and holes in EHP \cite{Butov1991}. The smallest density for EHP is determined by the exciton Mott transition $n_{\rm M} \sim 1/a_{\rm B}^2$ \cite{Schmitt-Rink1989}, where $a_{\rm B}$ is the exciton Bohr radius. For the CQW structures studied, $m_{\rm e} \simeq 0.07\,m_0$ and $m_{\rm h} \simeq 0.15\,m_0$ \cite{Lozovik2002}, the Bohr radius of the indirect excitons $a_{\rm B} \simeq 20$\,nm \cite{Dignam1991} and $n_{\rm M} \sim 1/a_{\rm B}^2 \sim 2 \times 10^{11}\,\mbox{cm}^{-2}$, so that the smallest linewidth for the EHP is $\Delta_{\rm EHP}^{\rm min} \simeq \pi \hbar^2 n_{\rm M} (1/m_{\rm e} + 1/m_{\rm e}) \sim 10$\,meV. In contrast, the linewidth of exciton emission can be well below this value. It is determined by the homogeneous and inhomogeneous broadening and is typically below 2\,meV in the CQW structures for the investigated
range of densities \cite{High2009}. The small emission linewidth $\lesssim 2$\,meV, which is characteristic for the inner ring reported in Refs.\,\cite{Butov2002, Ivanov2006, Stern2008}, and in the present paper, indicates that in all these experiments the inner PL ring forms in an exciton system rather than in EHP.

\section{Summary}

In summary, we studied kinetics of the inner ring in the exciton emission pattern. The formation time of the inner ring following the onset of the laser excitation is found to be about 30\,ns. The inner ring was also found to disappear within 4\,ns after the laser termination. The latter process is accompanied by a jump in the PL intensity. The spatial dependence of the PL-jump indicates that the excitons outside of the region of laser excitation, including the inner ring region, are efficiently cooled to the lattice temperature even during the laser excitation. The ring formation and disappearance are explained in terms of exciton transport and cooling.

\section{Acknowledgments}

We appreciate valuable discussions with R. Zimmermann. Support of this work by the ARO, NSF, EPSRC, and WIMCS is gratefully acknowledged.

\end{document}